\documentstyle[12pt,aps,epsf]{revtex}

\input{epsf}
\begin{document}

\title{The narrow peak on the frequency dependance of the susceptibility
of the harmonical oscillator in contact with thermostat with bosonic excitations.
Exactly-solvable model.}
\author{Kozlov G.G. {\it e-mail:} \hskip 4pt gkozlov@photonics.phys.spbu.ru}
\maketitle
\begin{abstract}

The relaxation of the harmonical oscillator, being in
contact with the thermostat whose
excitations can be treated as bosons is studied.  Temperatures
of the oscillator and thermostat
are supposed to be different at t=0.  An exact temporary
behavior of the inverse oscillator's
temperature is obtained. In low interaction limit the
temperature behavior is exponential.
When the interaction is increasing the temporary behavior
display oscillations.
{\it Narrow peak of the oscillator susceptibility is discovered.}
The exact  solution obtained  is compared with Bloch-Redfild
approximation.

\end{abstract}
\section{The problem description and results}

{\it If the reader knows some publications  related to the susceptibility
peak described below, please, send me the corresponding
reference.}

The problems of kinetics are known to be one of the most complicated
problems of physics. Existing approximate methods usually do not allow
one to estimate the quality of the approximation, and the plausibility
of the results is the main   criteria of the validity of the approximation
made. For this reason the simple exact-solvable models of dynamical
system and thermostat are of particular interest. The exact solution
can be compared in detailes with the solution, obtained by means of
the approximate method. This allow one to evaluate  used approximation
in a correct way. Moreover the exact solution can demonstrate
some new unexpected phenomena.
We consider the resonance of the susceptibility described below
to be one of them. In this paper we describe the model of this kind
with the harmonical oscillator as a dynamical system and system
having boson -like excitations as a thermostat. This
model was considered by Bogolubov \cite{Bog} for
classical vibrator. The typical example
of such thermostat is the phonon system of the crystal. The model's
Hamiltonian in frequency units has form:

\begin{equation}
H=H_p+H_s+H_{sp}
\end{equation}
$$
H_p=\sum_{q=1}^N \omega_q A_q^+A_q,\hskip 40pt   H_s=\omega_s B^+B
$$
$$
H_{sp}=C_0\sum_{q=1}^N(A_q B^+ +A_q^+ B),\hskip 40pt    C_0={b\over\sqrt N}
$$

Here $H_p$-- the Hamiltonian of the bosonic thermostat,
$\omega_q$ --the boson's dispersion (for the phonon thermostat -
the phonon frequencies)
$H_s$ -- The Hamiltonian of the harmonical oscillator,
$H_{sp}$-- the Hamiltonian of the interaction , $N$-- number of bosonic modes,
$C_0$ -- the interaction constant. The inverse proportionality of
this constant to the number of bosonic modes $N$ is typical
for all known interactions. $A_q^+ ,A_q (B^+, B)$ --the operators
of creation and annihilation  of bosones in q-mode (oscillator's energy
quantums). The following commutation rules are taking place:

\begin{equation}
[A_q,A_{q \prime}^+]=\delta_{q,q \prime}\hskip 2pt ,\hskip 20pt [B,B^+]=1
\end{equation}
We accept the simplest form for the bosonic spectrum:
\begin{equation}
\omega_q=\omega_s+{2W\over N}(q-N/2-3/4) \hskip 2pt,\hskip 20pt
q=1,...,N
\end{equation}
So, the oscillator frequency is coincide with bosones spectrum center.
Suppose the oscillator and thermostat has at $t=0$ the
temperatures $T_{p0}$ and $T_{s0}$  respectively.
So, the problem is to solve the Lioville equation for the
density matrix  $\rho$:

\begin{equation}
\imath \dot{\rho}=[H,\rho]
\end{equation}
with the initial condition:
\begin{equation}
\rho(0)=Z^{-1}\exp\bigg( -{H_p+H_{sp}\over T_{p0}} - {H_s\over T_{s0}}
\bigg),
\end{equation}
where
\begin{equation}
Z=\hbox{Sp} \exp\bigg( -{H_p+H_{sp}\over T_{p0}} - {H_s\over T_{s0}}  \bigg)
\end{equation}

Here we consider the temperature of  interaction equal to that
of thermostat. The unequilibrium state of this type can be
obtained in a following way. Let the oscillator's frequency
depends linearly on some external parameter $h$: $\omega_s=\bar{\omega}h$.
Let this  parameter has the value $h^\prime$  for a long time
and our system reached the equilibrium, having the density
matrix: $\sim\exp
-[\omega^\prime_s B^+B+H_{sp}+H_p]/T_{p0}$,
$\omega^\prime_s=\bar{\omega}h^\prime$. Now
let this parameter change instantaneously: $h^\prime \rightarrow h,$
and $h-h^\prime=\delta h $ and, therefore, $\omega^\prime_s \rightarrow \omega_s$.
After that the Hamiltonian of the system
has the form (1). It is easy to see
that right after jump of the parameter
$h$, the density matrix can be represented
in the form (5), with $T_{s0}=
(\omega_s /\omega_s^\prime)T_{p0}$.
Thermostat is supposed to be microscopic, so  all calculations below
 we perform in thermodynamic limit $N\rightarrow\infty$. Lastly note that
 the above model can describe the spin relaxation at temperatures low
 enough as compared with spin levels splitting in
 magnetic field.  In this case the parameter $h$ can be associated
 with magnetic field.

Thus problem is described, we now itemize the main points of the
solution. For this reason we make the following general remark.
Consider the set of operators $O_i\hskip 2pt,\hskip 10pt i=1,... M$
to be closed with respect to the commutation with Hamiltonian
in the following sense:

\begin{equation}
[H,O_i]=\sum_{k=1}^M\alpha_{ik}O_k,
\end{equation}
where $\alpha_{ik}$ - numbers.
Take now the density matrix in the form:
\begin{equation}
\sigma (t)=\sum_{i=1}^M \beta_i(t)O_i,
\end{equation}
substitute it in (4). Making use of (7), we calculate commutators
and equalize the factors corresponding to the similar operators.
We obtain the following set of equations for $\beta_i(t)$ functions:

\begin{equation}
\imath \dot{\beta_i}=\sum_{k=1}^M \beta_k \alpha_{ki}\hskip 2pt,\hskip 10pt i=1,...,M
\end{equation}
 After solving this set of equations we obtain the solution of Lioville
equation in the form (8).
Note now that the set of operators $A_q^+A_q ,\hskip 3pt A_qB^+,
\hskip 3pt A_q^+B,\hskip 3pt q=1,...,N$ and $BB^+$ is closed in
the above sense with respect to the commutation with Hamiltonian (1)
and we can use the above  method as follows.
Let us put into consideration the auxiliary density matrix $\sigma$:

\begin{equation}
\sigma=\sum_{q=1}^N \bigg(\beta_q A_q^+A_q + \gamma_qA_q B^+ + \gamma_q^*A_q^+ B
\bigg)+\beta_s B^+B
\end{equation}

Substituting it in (4) and equalizing the factors corresponding
to the similar operators we obtain the following set of equations:

\begin{equation}
\imath\dot{\beta_q}=C_0(\gamma_q-\gamma_q^*)
\end{equation}
$$
\imath\dot{\beta_s}=-C_0\sum_{q=1}^N(\gamma_q-\gamma_q^*)
$$
$$
\imath\dot{\gamma_q}=(\omega_s-\omega_q)\gamma_q+C_0(\beta_q-\beta_s)
$$
$$
\imath\dot{\gamma_q^*}=-(\omega_s-\omega_q)\gamma_q^*-C_0(\beta_q-\beta_s)
$$

We solve this system with initial conditions:

\begin{equation}
\beta_q(0)=\omega_q/T_{p0},\hskip 6pt \beta_s(0)=\omega_s/T_{s0},\hskip 6pt
\gamma_q(0)=\gamma_q^*(0)=C_0/T_{p0}
\end{equation}

After that we use known property of Lioville equation \cite{Al} (p. 170) :
if matrix $\sigma$ satisfy the Lioville's equation, then the arbitrary
analitical function of this matrix satisfy the Lioville's equation too.
Taking this into account, it is easy to see that matrix
$\rho=Z^{-1}\exp(-\sigma)$ satisfy the Lioville's equation
 with desired initial conditions (5).
 The final solution of the problem we obtain by turning to the
thermodynamic limit $N\rightarrow\infty$.
 It is seen that functions $\beta_q,\beta_s$ can be called the "dimensionless inverse
temperatures" of the $q$  bosonic mode and dynamical system respectively.
We call  $\gamma_q$ functions the "dimensionless inverse temperatures of the
interaction" with the bosonic mode $q$.
An exact solution of  (11) performed in the second section lead to the
following formula for the temporary dependance of the inverse
temperature of the oscillator $T_{s}^{-1}=\beta_s(t)/\omega_s$ :

\begin{equation}
{1\over T_s}\bigg(t\bigg)={1\over T_{s0}}+\bigg({1\over T_{p0}}- {1\over T_{s0}}  \bigg)
{2\tau\over\pi}\int_0^W {1-\cos\Omega t \over 1+\bigg(\Omega\tau-\Delta(\Omega)\bigg)^2}d\Omega
\end{equation}
\begin{equation}
\tau^{-1}={\pi b^2\over W }= 2\pi b^2 \rho_0
\end{equation}
\begin{equation}
\Delta(\Omega)={1\over\pi}\ln{W+\Omega \over W-\Omega}
\end{equation}

Here $\rho_0\equiv 1/2W$ - is normalised to unit density of bosons modes.
If $b/W<<1$ one can extend the integration up to the infinity and drop
 the logarithmic term $\Delta(\Omega)$ in denominator.
After that the integral can be calculated and we obtain:

\begin{equation}
{1\over T_s}\bigg(t\bigg)={1\over T_{s0}}+\bigg({1\over T_{p0}}- {1\over T_{s0}}  \bigg)
\bigg( 1-\exp -\bigg|{t\over\tau}\bigg|\bigg)
\end{equation}

The temporary dependances, calculated by formula (13) for different $b/W$ ,
are presented on fig.1. It is seen that when interaction is small enough,
the kinetics is exponential. Increasing of the interaction is
accompanied by occurrence of oscillatory kinetics.
We present here the expression for the inverse temperature of the bosonic q- mode
 $q\hskip 30pt$ $1/T_q=\beta_q/ \omega_q$:

\begin{equation}
{1\over T_q} = {1 \over T_{p0}}-
\end{equation}
$$
\bigg({1\over T_{p0}}- {1\over T_{s0}}  \bigg)
\bigg({2 b^2 \omega_s \tau \over N \omega_q}  \bigg)
\bigg({2 \over \pi} \int_0^W {(1-\cos \Omega t)d\Omega \over (\Omega^2-\omega^2)[1+(\Omega\tau-\Delta(\Omega))^2]}
+ (1-\cos \omega t){\tau-\Delta(\omega) / \omega \over 1+(\omega \tau-\Delta(\omega))^2}\bigg)
$$
$$
\omega^2 \equiv (\omega_q - \omega_s)^2 + 2C_0^2
$$

If $b/W<<1$ this expression can be simplified:

 \begin{equation}
 {1\over T_q}={1 \over T_{p0}}-
 \end{equation}
$$
\bigg({1\over T_{p0}}- {1\over T_{s0}}  \bigg)
\bigg[{2b^2\tau\omega_s\over N\omega_q(1+\omega^2\tau^2)}  \bigg]
\bigg[\tau(1-\cos\omega t)+{\sin\omega |t| \over\omega}-\tau\bigg(1-\exp - \bigg | {t\over\tau}\bigg |\bigg)\bigg]
$$

Formula (14) can be rewritten in the form:

\begin{equation}
\tau^{-1}=2\pi b \rho_0(\omega_s)
 \end{equation}

$\rho_0(\omega_s)$ - density of the resonance bosonic modes.

Let us now return to the dropped logarithmic term which can be of
particular interest for the following reason.
The above  problem describes the 'system's response' on the jump of
the parameter $h$ .
 If we consider the 'system's response' as $\delta (1/T_s)\equiv
(1/T_s-1/T_{p0})$, then $\delta (1/T_s)=\Phi(t)\delta h$,
where $\Phi(t)$ can be derived from (13).
Let parameter $h$ gains small harmonical increment $\delta h_\omega \exp\imath\omega
t$. The corresponding small modulation $\delta (1/T_s)_\omega\exp\imath\omega t$
can be expressed in terms of susceptibility:
 $$
 \delta \bigg({1\over T_s} \bigg)_\omega=\chi_\omega \delta
 h_\omega
 $$
On the other hand, the linear response theory \cite{Al} (ð. 88) lead to the following
relationship between the susceptibility and 'system's response'
on the jump $\Phi(t)$:
$$
\Phi(t)= {1\over 2\pi}\int_{-\infty}^{+\infty}{\chi_\omega e^{\imath\omega t} \over
\imath\omega+0}\hskip 2pt d\omega \hskip 2pt,\hskip 40pt
\chi_\omega=\imath\omega\int_0^{+\infty} e^{-\imath\omega t}\Phi(t) dt
$$
The imaginary part of susceptibility can be explicitly calculated.
It  differs from zero only when $-W<\omega<W$ and is equal to:

$$
\hbox{Im}\hskip 2pt \chi_\omega=\chi_{\infty}
\hskip 3pt{\omega\tau\over
1+(\omega\tau-\Delta(\omega))^2}\hskip 30pt
\chi_{\infty}\equiv -{\bar{\omega_s}\over T_{p0}\hskip
2pt\omega_s}
$$
It is seen that susceptibility  Im$\chi_\omega$ exhibit two
extraordinary extremums whose positions  $\pm\omega_0$ (when
 $b/W<<1$)  can be defined from the equation:
$$
\omega_0\tau=\Delta(\omega_0)
$$

It is easy to see that $\hskip 10pt\omega_0\approx W-0$.
So, if  $\omega=\omega_0$ the modulation of oscillator's inverse temperature in
the orthogonal phase (with respect to $\delta h$), is $W\tau=(W/b)^2/\pi>>1$
 times greater than the adiabatic modulation $\chi_{\infty}\delta h$ .
If interaction is small enough, it is possible to estimate
the absorption under the action of modulation
$\delta h=\delta h_\omega\sin\hskip 2pt \omega t$.
The rate of energy changing  $dE/dt$ for whole system (oscillator+thermostat)
can be calculated as:

$$
{dE\over dt}={d\over dt}\hbox{Sp}\hskip 2pt \rho H=\hbox{Sp}\hskip 2pt \rho {d\over
dt}H_s=
\bar{\omega_s}\hskip 2pt\dot{\delta h}\hskip 4pt\hbox{Sp}\hskip 2pt\rho B^+B=
\bar{\omega_s}\dot{\delta h}\hskip 4pt<n>
$$

If the interaction is small,  terms $\sim\gamma_q$ can be omitted
in  $\rho\sim\exp(-\sigma)$.
In this case the average number of quantums $<n>$ is defined by
 the Bose-Einstein function $<n>=1/(e^{\beta_s}-1)$ with $\beta_s=
\omega_s/T_s$  depending on time.
Linear response $\delta\beta_s$ related to  the modulation
$\delta h$ can be determined by above expression for susceptibility.
Furthermore $dE/dt$  should be averaged over the period of oscillations.
For this reason only the component of $\delta (1/T_s)$ having the same phase as
$\dot{\delta h}$ is important.
It is defined by above expression for the imaginary part of the susceptibility.
Taking all this into account it is easy to
see that $\bigg\langle {dE/ dt}\bigg\rangle$ is differes
from zero only if $0<\omega<W$ and in this frequency interval can be calculated as:

$$
\bigg\langle {dE\over dt}\bigg\rangle=
{\delta\omega_s^2\over 2 T_{p0}}\hskip 3pt {e^\beta\over (e^\beta-1)^2}\hskip 3pt
{\omega^2\tau\over 1+[\omega\tau-\Delta(\omega)]^2}\hskip 30pt
\beta\equiv {\omega_s\over T_{p0}},\hskip 5pt
\delta\omega_s\equiv\bar{\omega_s}\delta h_\omega
$$

The frequency behaviour of the absorption is presented diagrammatically on fig.1 (c).
It is seen that the absorption  exhibit narrow maximum at $\omega=\omega_0$.
The occurrence of this maximum is somewhat  unexpected. This peak do not depend
on details of the bosonic spectrum.

\section{The equation for the inverse temperatures}

In this section we solve the set of equations (11) which can
be rewritten as:
\begin{equation}
\ddot{I_q}=-[(\omega_q-\omega_s)^2+2 C_0^2]
I_q-2C_0^2\sum_{q\prime=1}^N I_{q\prime}
\hskip 30pt q=1,...,N
\end{equation}
$$
I_q\equiv\gamma_q^*-\gamma_q
$$

As a starting point we find the solutions of this equation
having the harmonical temporary dependance - normal modes
$ e_q ^{\Omega}\exp( \imath\Omega t),\hskip 10pt
q=1,...,N$.

After substituting in (20) we obtain the following equations
for the amplitudes $ e_q ^{\Omega}$:

\begin{equation}
e_q^{\Omega}=-{2C_0^2
S\over (\omega_s-\omega_q)^2+2C_0^2-\Omega^2}
\hskip 30pt q=1,...,N
\end{equation}
where
\begin{equation}
S\equiv\sum_{q=1}^N e_q^\Omega
\end{equation}
Substituting (21) in (22) we obtain for squares of eigen frequences
$\Omega^2$:

\begin{equation}
\sum_{q=1}^N {2C_0^2\over
(\omega_s-\omega_q)^2+2C_0^2-\Omega^2}=-1
\end{equation}

For the bosonic spectrum in the form (3) this equation have $N$
undegenerated roots.
Using  (21) we obtain for the normal modes the following expression:

\begin{equation}
e_q^\Omega={N_0(\Omega)\over
(\omega_s-\omega_q)^2+2C_0^2-\Omega^2}
\end{equation}

Where $N_0(\Omega)$ is defined as:
\begin{equation}
N_0^{-2}(\Omega)=\sum_{q=1}^N \bigg[{1\over (\omega_s-\omega_q)^2+2C_0^2-\Omega^2}
\bigg]^2
\end{equation}

(24) is the eigen mode of (20) only if  $\Omega^2$ is
coincide with one of the roots of (23).
The initial conditions for quantities $I_q$ can be obtained from (12):

\begin{equation}
I_q(0)=0\hskip 2pt,\hskip 30pt \dot{I_q(0)}=2\imath
C_0\omega_s\bigg({1\over T_{p0}}-{1\over T_{s0}} \bigg)
\end{equation}

Introducing the solution of (20) in the form of expansion over the
normal modes, we can write down the solution of (20) as:

\begin{equation}
I_q(t)=2\imath
C_0\omega_s\bigg({1\over T_{p0}}-{1\over T_{s0}} \bigg)
\sum_{\Omega q\prime}\hskip 3pt {e_q^\Omega \hskip 3pt e_{q\prime}^\Omega \hskip 3pt\sin \Omega
t\over \Omega}
\end{equation}

$\Omega$ runs over the set of eigen frequences obtained from (23).
From (23) and (24) one can obtain:

\begin{equation}
\sum_q\hskip 3pt e_q^\Omega \hskip 3pt =-{N_0(\Omega)\over 2C_0^2}
\end{equation}

Using this relationship we obtain the solution of (20) in the form:

\begin{equation}
I_q=-\imath {\omega_s\over C_0}\bigg({1\over T_{p0}}-{1\over
T_{s0}}\bigg)
\sum_\Omega {N_0^2(\Omega)\hskip 3pt\sin \Omega t \over \Omega \hskip 3pt[(\omega_q-\omega_s)^2+
2C_0^2-\Omega^2]}
\end{equation}
Taking into account (28), (27), (11)  and initial conditions
for $1/T_s=\beta_s/\omega_s$ it is easy to obtain:

 \begin{equation}
\bigg( {1\over T_s}\bigg)=\bigg( {1\over T_{s0}}\bigg)+{1\over 2C_0^2}\bigg({1\over T_{p0}}
-{1\over T_{s0}}\bigg)
\sum_\Omega N_0^2(\Omega)\hskip 5pt {1-\cos\hskip 3pt \Omega t \over \Omega^2}
\end{equation}

The sum in this formula can be replaced by integral if
density of squares of eigen frequences $\Omega^2$ is known.
We calculate this density as follows. The above eigen
frequences arises in the following problem for eigen values:
$\hat{M} {\bf e}=\Omega^2{\bf
e}$, where matrix $\hat{M}$ is defined as:

\begin{equation}
\hat{M}=\hat{M}_0+\hat{V}
\end{equation}
$$
\hat{M}_0=\left(\matrix{
d_1 & \cdots & 0   \cr \vdots & \ddots & \vdots  \cr 0 & \cdots &  d_N
}\right )
\hskip 30pt \hat{V}=
2C_0^2
\left(\matrix{
1 & \cdots & 1 \cr \vdots & \ddots & \vdots  \cr 1 & \cdots &  1
}\right )
$$

$$
d_m=(\omega_s-\omega_q)^2+2C_0^2 \rightarrow
{W^2\over N^2}(m+1/2)^2+2C_0^2
$$

The diagonal elements $d_m$ are written for the
bosonic spectrum in the form (3).
Let us put into consideration the Green's function:

\begin{equation}
\hat{G}(E)=(E-\hat{M})^{-1}
\end{equation}

Then the density of squares of eigen frequences of interest $ D(\Omega^2)$
is defined by known formula:

\begin{equation}
D(E)= -{1\over \pi} \hskip 3pt \hbox{Im Sp}\hskip 3pt
\hat{G}(E),\hskip 33pt E=\Omega^2+\imath\delta \hskip 30pt
\delta\rightarrow +0
\end{equation}

Considering $\hat{V}$  as perturbation for  $\hat{M}_0$, one can write down
the Dyson's expansion for Green's function:

\begin{equation}
\hat{G}=\hat{G}_0+\hat{G}_0\hat{V}\hat{G}_0
+\hat{G}_0\hat{V}\hat{G}_0\hat{V}\hat{G}_0+...
\hskip 30pt \hat{G}_0\equiv (E-\hat{M}_0)^{-1}
\end{equation}

This series can be exactly summed and for the Sp $\hat{G}$ of
interest we obtain:

\begin{equation}
\hbox{Sp}\hskip 2pt \hat{G}(E)={\Gamma(E)\over 2C_0^2}-{ \partial \Gamma/\partial E\over
1-\Gamma(E) }
\end{equation}
where
\begin{equation}
\Gamma(E)\equiv 2C_0^2\sum_m {1\over E-d_m}=
\hskip 3pt2C_0^2\hskip 3pt\hbox{Sp}\hskip 3pt \hat{G_0}
\end{equation}

For the bosonic spectrum in the form (3) in the thermodynamic
limit one can write down the explicit expression for $\Gamma(E)$:

\begin{equation}
\Gamma(E)=\bigg({2b^2\over N}\bigg)\sum_{m=0}^N{1\over
E-(W/N)^2(m+1/2)^2-2b^2/N}=
\end{equation}
$$
=2b^2\int_0^1{d \xi \over E-(W\xi)^2}=
{b^2\over W \sqrt E}\hskip 4pt\ln {\sqrt E +W \over \sqrt E - W}
$$

 Factor (25) can be presented in terms of $\Gamma(E)$ as:

\begin{equation}
N_0^2(\Omega)=-{2C_0^2\over \partial\Gamma/\partial E}\hskip 30pt
E=\Omega^2
\end{equation}

And taking into account (33), (35) we obtain the following relationship:
\begin{equation}
{1\over 2C_0^2}\sum_\Omega {N_0^2(\Omega)\over\Omega^2}\bigg(
1-\cos\hskip 3pt \Omega t\bigg)=-{1\over\pi}\hbox{Im}\int{dE\over
1-\Gamma(E)}\hskip 3pt{1-\cos\hskip 3pt \sqrt E t\over E}
\end{equation}

Making use of (30), (39) it is now easy to obtain (13).
Formula (17) can be obtained in the same manner.

\section{The comparison with Bloch-Redfild's theory.}

This theory \cite{Al} (ð. 60) lead to the following kinetic equation for
the diagonal elements of density matrix of above oscillator:

\begin{equation}
\dot{\rho_n}=\rho_{n+1} W_{n+1 \rightarrow n} +\rho_{n-1} W_{n-1 \rightarrow
n}-\rho_n (W_{n \rightarrow n+1}+ W_{n \rightarrow n-1})
\end{equation}
where $ W_{n \rightarrow n \pm 1}$ :
\begin{equation}
 W_{n \rightarrow n+1}={1\over \tau}\hskip 4pt {n+1\over \exp(\omega_s/T_{p0})-1}
\end{equation}
$$
W_{n+1 \rightarrow n}=W_{n \rightarrow n+1}\hskip 3pt \exp\hskip
2pt \big(\omega_s/T_{p0}\big)
$$

With $\tau$ determined by formula (14).
It is seen that the temporary behaviour of density matrix in
Bloch-Redfild approximation do not agreed with above exact
solution. But the relaxation time of  the average energy of the
oscillator $<n>=\sum_n\rho_n n$ is coincide with (14)
This can be shown by multiplying (40) by $n$ and summing over all states:

\begin{equation}
\dot{<n>}={1\over\tau}\hskip 2pt \bigg[ {1\over \exp (\omega_s/T_{p0})-1}-<n>\bigg]
\end{equation}

\begin{figure}
\epsfxsize=400pt
\epsffile {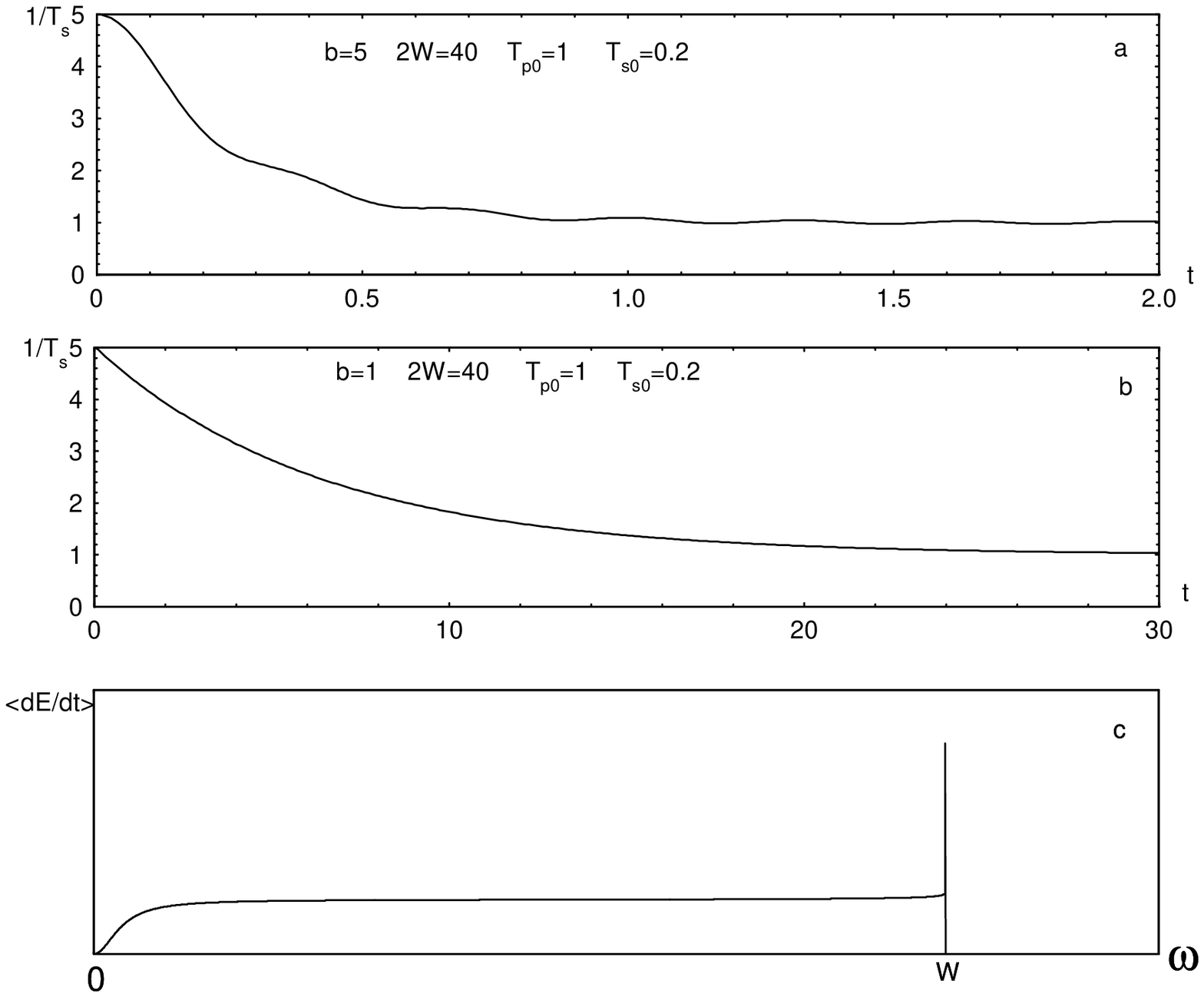}
\caption{ The kinetics of the inverse temperature of the oscillator for
various strength of interaction (a,b). The frequency
dependance of absorption (c).}

\end{figure}

\end{document}